\documentclass[prd,nofootinbib,preprintnumbers,floatfix]{revtex4}
\usepackage{graphicx}
\usepackage{natbib}
\newcommand{\rthis}[1]{\textcolor{black}{#1}}
\usepackage[plainpages=false, colorlinks=true, anchorcolor=blue, linkcolor=blue, citecolor=blue, bookmarks=false]{hyperref}
\begin{document}

\newcommand{\physrep}{Physics Reports}
\newcommand{\ssr}{Space Science Reviews}
\newcommand{\apjl}{Astrophys. J. Lett.}
\newcommand{\apjs}{Astrophys. J. Suppl. Ser.}
\newcommand{\aap}{Astron. \& Astrophys.}
\newcommand{\aapr}{Astronomy \& Astrophysics Review}
\newcommand{\aj}{Astron. J.}
\newcommand{\araa}{Ann. Rev. Astron. Astrophys. } %ARA$\&$A}
\newcommand{\mnras}{Mon. Not. R. Astron. Soc.}
\newcommand{\apss} {Astrophys. and Space Science}
\newcommand{\jcap}{JCAP}
\newcommand{\pasj}{PASJ}
\newcommand{\pasp}{PASP}
\newcommand{\pasa}{Pub. Astro. Soc. Aust.}
%\usepackage[mathlines]{lineno}% Enable numbering of text and display math
%\linenumbers\relax % Commence numbering lines

%\usepackage[showframe,%Uncomment any one of the following lines to test 
%%scale=0.7, marginratio={1:1, 2:3}, ignoreall,% default settings
%%text={7in,10in},centering,
%%margin=1.5in,
%%total={6.5in,8.75in}, top=1.2in, left=0.9in, includefoot,
%%height=10in,a5paper,hmargin={3cm,0.8in},
%]{geometry}

%\preprint{APS/123-QED}

\title{A test of spatial coincidence between CHIME FRBs and IceCube  TeV  energy neutrinos }% Force line breaks with 

\author{Shantanu \surname{Desai}}%
\altaffiliation{E-mail: shntn05@gmail.com}

\date{\today}
\affiliation{Department of Physics, IIT Hyderabad, Kandi, Telangana-502284, India}

\vspace{10pt}

%\affiliation{$^{2}$ Department of Physics, Indian Institute of Science Education and Research Bhopal, Bhopal 462066, India}

\begin{abstract}
We  search for a spatial association between the CHIME FRBs and  IceCube neutrinos detected in the TeV energy range,   by  counting the total number of  neutrino-FRB pairs with  angular separations of $< 3 $ degrees, as well as within the observed neutrino error circle. This  number constitutes the total signal events, which is  then compared to the total background, corresponding to the null hypothesis of no spatial association.  The background   was obtained  from the  total neutrino-FRB matches  in off-source angular windows with the same solid angle as the signal window. We do not find any statistically significant excess compared to the background. Therefore, we conclude that there is no evidence for  an angular correlation between the IceCube neutrinos in the TeV energy range and CHIME FRBs. For each of these searches, we report 90\% Bayesian credible interval upper limits on the observed FRB-induced neutrinos.
\end{abstract}

\maketitle

\section{Introduction}
We search for a statistically  significant  angular correlation between the Fast-Radio bursts (FRBs) detected by the Canadian Hydrogen Mapping Intensity experiment (CHIME)~\cite{Chime} and  neutrinos detected by the IceCube \rthis{Neutrino Observatory} in the energy range between 0.1 to 3 TeV~\cite{icecube}.  
FRBs are short-duration radio bursts, which were  discovered in 2007~\cite{Lorimer07,Petroff21}. Although the exact nature of FRBs is still unknown (see ~\cite{Platts} for a compendium of all FRB models),  the recent association between an FRB and a galactic magnetar SGR J1935+2154~\cite{CHIMEmagnetar}, has put magnetars as the forefront candidates for FRBs. 

%The statistical significance of the aforementioned correlation in L21  was found to be 21.3$\sigma$.  This search was carried out by looking for spatial coincidences between the FRBs and neutrinos within the error circle  of each neutrino.  A total of 192900 such matches were found.
%This was then compared with the background, which was estimated by generating synthetic FRB sources randomly distributed  within the observed RA,DEC range, and  doing the same angular search on each set of these synthetic catalogs. The mean number of matches and its standard deviation was thus  estimated to be $126030 \pm 3056$. Comparing this to the observed matches, yields a detection significance of 21.3$\sigma$. The significance is  maximum between the energy range of 0.3 -1 TeV. L21 then pinned the observed excess to 20 non-repeater FRBs. They also showed that these 20 FRBs are representative of the entire CHIME sample.

Since the origin of the diffuse astrophysical neutrino flux detected by IceCube~\cite{IceCube13,IceCubescience,IceCube14}  is still a mystery~\cite{Meszaros17,Halzen}, it would be interesting to see  if FRBs contribute to the diffuse astrophysical neutrino flux, given the surplus of new FRBs discovered by CHIME.
Note however that previous \rthis{ spatial  and} temporal  searches  for neutrino emission  from FRBs (detected prior to CHIME) using IceCube, have found null results~\cite{Fahey,icecube2018,IcecubeFRB}.
Furthermore, null results have also been reported in searches for  neutrinos associated with  magnetars from 
both IceCube and Super-Kamiokande~\cite{Icecubemagnetar,sknuastro,showering,Desaithesis}.  

Therefore, we set out to do this  search using the first CHIME FRB catalog,  using an independent method of estimating the background by counting the observed matches in the off-source regions, \rthis{where by ``off-source'' we refer to angular directions far from the putative astrophysical source, wherein  any observed neutrinos would be random coincidences and only due to the background.} This search is similar to searches for dark matter WIMPs and astrophysical point sources with the Super-Kamiokande experiment ~\cite{superkgrb,superkwimp,sknuastro,showering}.  One difference between this search and all other IceCube-based FRB searches is that the previous works carried out both temporal \rthis{and spatial} searches from each of the FRBs.  Here, we carry out a stacked search from all the observed FRBs without imposing any time-window around the FRBs. The \rthis{first CHIME catalog published in Ref.~\cite{Chime} (which we have used for our analysis)  contains 18 repeating FRBs}. However, the actual number  of FRBs which emit bursts multiple times could be larger in case the radio  emission happens outside the CHIME duty cycle or when the source is not within the CHIME field of view. Secondly, although hadronic acceleration has been proposed in the regions of the progenitors, there is considerable uncertainty between the radio emission detected in FRBs and any putative neutrino emission~\cite{Li14,Patrick}. \rthis{ Therefore, in this work we only look for spatial coincidences. Furthermore,  one practical consideration for carrying out only spatial searches is that currently there is no temporal overlap between the two datasets. In the future, we shall extend this analysis by also carrying out a temporal search.}   

This manuscript is structured as follows.
The dataset used for our analysis is described in Section~\ref{sec:dataset}. Our analysis and results can be found in Sect.~\ref{sec:results}. We conclude in section~\ref{sec:conclusions}.

\section{Datasets}
\label{sec:dataset}
The public IceCube  point source catalog contains 1,134,450 neutrinos spanning a ten year time-span from 6th April  2008 to 8 July 2018~\cite{icecube,icecubedata}. Each neutrino event is provided with an associated \rthis{right ascension (RA), declination (DEC)},  \rthis{reconstructed muon} energy, angular error, and finally the detector zenith and azimuth angle. \rthis{The reconstructed muon energy provides a   lower limit of the parent neutrino energy.} The \rthis{muon} energy spans almost nine decades from 10 GeV to approximately 30 PeV. The average angular resolution is $0.85^{\circ}$. The neutrino positions span both the hemispheres.
The CHIME FRB catalog contains 491 FRBs, of which 18 are repeaters. These bursts have been detected between between 25th July 2018 to 2019 July 1
and measured from 400-800 MHz. \rthis{Therefore, there is no temporal overlap between the public IceCube catalog and the CHIME FRB dataset.}
The CHIME collaboration provides 54 attributes for each FRB. The average  error in RA and DEC for the CHIME FRBs  is about 0.16 and 0.18 degrees, respectively. The FRB catalog mostly covers the Northern Hemisphere and only contains eight \rthis{FRBs} with DEC$<0^{\circ}$. More details on the CHIME FRB catalog can be found in Ref.~\cite{Chime}.
We do not make a cut on the declination of the IceCube neutrinos, as we would like to use all of the observed neutrinos for background estimation. We only include IceCube events with angular resolution $< 3^{\circ}$. This gives a total of 1,100,993 events, which we use for our analysis.
\rthis{The  $3^{\circ}$ cut is a conservative choice and is  based on a similar study done in Super-K astrophysical neutrinos searches~\cite{sknuastro}. With this cut we include 97\% of the full IceCube dataset.}
Note that  we do not consider the arrival time of the neutrino or the time of the FRB burst, since we are only doing a spatial analysis.  

\section{Analysis and Results}
\label{sec:results}

\subsection{Coincident searches within $3^{\circ}$}
If there is a statistically significant association between the CHIME FRBs and IceCube neutrinos, one should see an  excess at small angular (on-source) separations between the FRBs and neutrinos, after stacking all the pairs. Since we do not have redshift measurements for the FRBs, we assume equal weights for each FRB. The total  number of pairs within the on-source window should be much greater  than the background, which can be obtained by considering the observed matches in the off-source regions at large angular separations. Any excess in the on-source region compared to the background can be regarded as being of astrophysical  origin, and can be attributed to FRBs in this case. 

We calculate the angular separations between all the FRBs and observed neutrinos, and bin the number of coincident pairs into bins of $3^{\circ}$, as this is the maximum uncertainty in the neutrino direction, because of the cut we have used. At TeV energies, any coincidences between the FRBs and neutrinos at angular separations greater than $3^{\circ}$ are chance coincidences and can be considered as background for our search.  \rthis{We do this analysis separately for  four energy intervals: 0.1-3 TeV, 1-3 TeV, 3-10 TeV,  and finally considering the full dataset.}

\begin{figure}
\centering
    \includegraphics[scale=0.7]{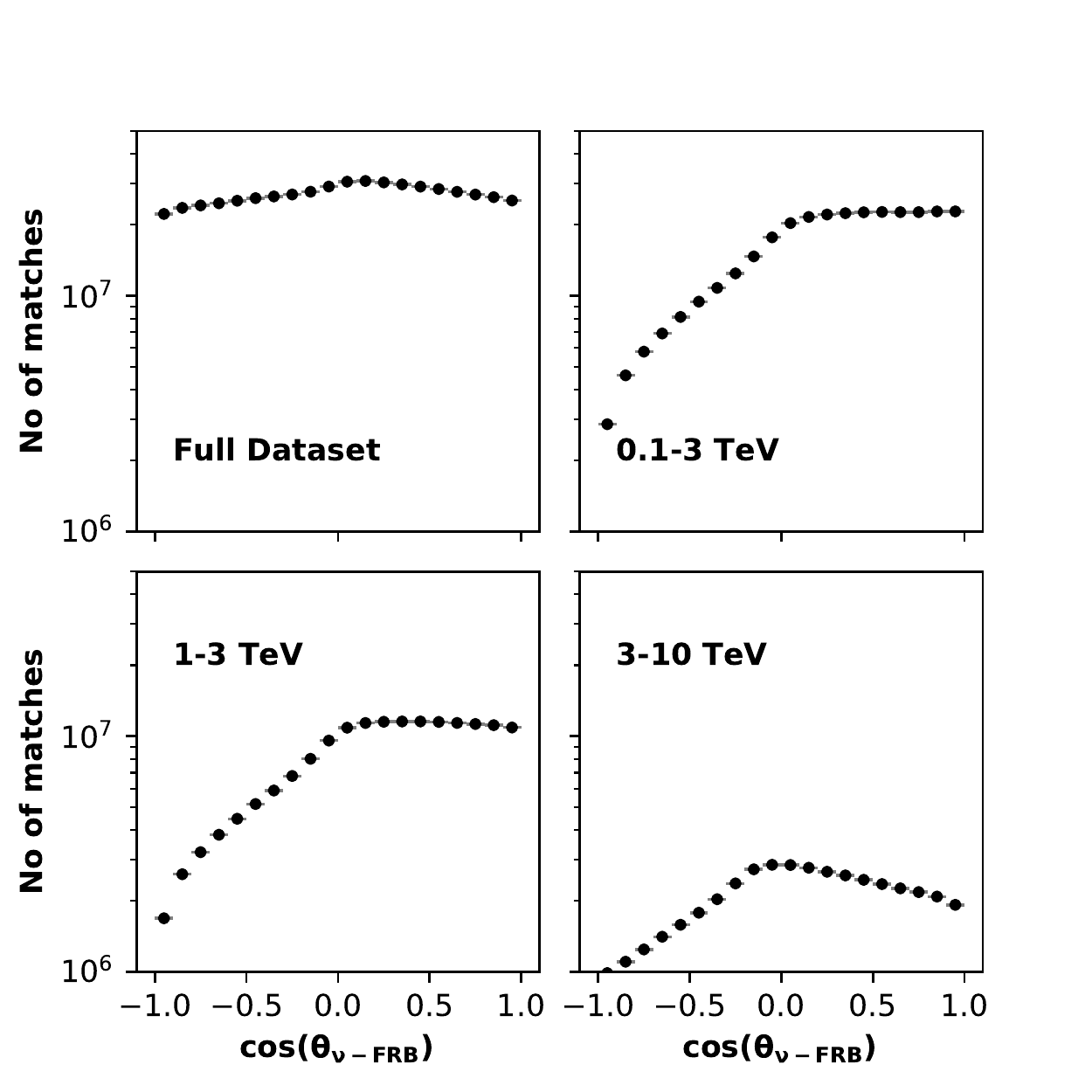}
    \caption{The  angular distribution between of all IceCube neutrinos with respect to the  CHIME FRBs for four different \rthis{muon} energy intervals, binned in $\cos (\theta)$ bins of width 0.1.}
\label{fig1}.
\end{figure}

We first show the full angular  distribution  between all the FRBs and neutrinos, by binning the data into cosine angle bins of width equal to  0.1.  \rthis{This is an exploratory test}, as for  any strong angular correlation, one should see a peak in the full angular distribution at close to $0^{\circ}$ separations,
similar to the  peak observed in solar neutrinos, near the direction of the Sun~\cite{sksolar}. This distribution can be found in Fig.~\ref{fig1}, where the data is binned in cos($\theta$) bins of size equal to 0.1, where $\theta$ is the angular separation between each pair of  FRBs and neutrinos.  This distribution is also shown for the full dataset as well as three different equidistant (in log space) energy intervals   from 0.1-10 TeV. 
In the case of a  statistically significant association, one should see an excess in the $\cos(\theta)$ bin between 0.9-1 (similar to that seen in solar neutrinos~\cite{sksolar}). We do not see such an excess. \rthis{We however note that the zenith angle distribution is not uniform for some of the energy bins. This is due to a combination of multiple factors. Almost all the FRBs are located in the northern celestial hemisphere. Although, the full IceCube dataset is spread over both the hemispheres,
almost all the neutrinos with reconstructed muon energies between 0.1-3 TeV are located in the northern hemisphere. Therefore, there are very few events in the antipodal directions of the FRBs. Therefore, $\cos(\theta)$ shows a downward trend for the second and third panels, as we approach $\cos(\theta)=-1$. Finally,  most of  the IceCube events with energies between 3 and 10 TeV show a pile-up near the  declination of $0^{\circ}$, with a steeply falling distribution towards $90^{\circ}$.  This is then reflected in the zenith angle distribution, which thereby falls off as $\cos(\theta)$ approaches one.}

%We then show a zoomed version of this figure, where we plot the stacked angular separation between the neutrinos and FRBs  between 3 and 30 degrees, similar to WIMP searches done in Super-K~\cite{superkwimp}.
%Since we have culled all neutrino events with error $>3^{\circ}$, all neutrinos associated with FRBs should be detected within a cone of $3^{\circ}$. For any statistically significant association between the FRBs and neutrinos, the number of matches within $3^{\circ}$ should be much larger  than the background. This plot can be found in Fig.~\ref{fig2}. Again, by eye we do not see any excess within the $3^{\circ}$ bin.

To estimate the background, we bin the cosine of the angular  between the neutrino-FRB patches  at separations less than $7.35^{\circ}$ (corresponding to the angular \rthis{separation} $\cos \theta \in$ [0.9,1]) into six equidistant (in $\cos \theta$) bins. This ensures that all the background bins have the same solid angle as the bin containing the signal events. Note that this procedure would emulate the background due to atmospheric neutrinos as well as any non-FRB induced  astrophysical neutrinos.  This aforementioned plot is shown in Fig.~\ref{fig3}. The right-most bin corresponds to an angular separation of $3^{\circ}$, which is where all the signal events would be seen.  The background is obtained by averaging the total number of events in the remaining five bins, which correspond to the off-source region.  From Fig.~\ref{fig3}, we do not see any statistically significant excess in the 0-3 degree bin,
compared to the off-source bins.
Therefore,  there is no statistically significant association between the IceCube neutrinos and the CHIME FRBs.

In order to directly compare the observed signal and background, we tabulate the observed signal events and compare them to the expected background obtained, after averaging all the coincident matches in the off-source bins. These results are summarized in Table~\ref{tab:my_label}.
The error in the background corresponds to the Poisson error. As we can see, for all the energy intervals, the number of signal events is smaller or comparable to the background. \rthis{The only case when the total number of signal events is greater than the background is for the energy range 3-10 TeV where we observe 25135 signal events compared to an estimated background of 25021 events. However this excess is less than $1\sigma$. In order to claim a $3\sigma$ detection, we should have seen  at least
25495 coincident pairs.}

Therefore, from this analysis, we do not find any statistically significant  excess between the IceCube neutrinos and CHIME FRBs, when  compared to the  background.   We than calculate 90\% credible interval flux limits using Bayesian  analyses using the prescription in 2021 PDG~\cite{PDG} for for positive definite signal events corresponding to a Poisson distribution, which can be obtained as follows:
\begin{equation}
     \frac{\int \limits_0^{s_{up}} P(N|s+b) ds}{\int \limits_0^{\infty} P(N|s+b) ds} = 0.9,
     \label{eq:upperlimit}
\end{equation}
Here, $s_{up}$ is the desired 90\% upper limit, and $P(N|s+b)$ is the Poisson probability mass function for $N$ observed events with mean given by $b+s$, where $b$ is the observed background. These 90\% credible interval upper limits can be found in the last column in Table~\ref{tab:my_label}.

\begin{table}[]
    \centering
    \begin{tabular}{|c|c|c|c|}
    \hline
    \textbf{Muon Energy range} & \textbf{Signal}  & \textbf{Background} & \textbf{90\% credible flux limit}\\ 
    \hline
    
    Full dataset & 339122 & $340146 \pm 583$ & 525 \\ \hline
    0.1 -3 TeV & 310012 & $310725 \pm 557$ & 583  \\ \hline
    1-3 TeV & 146762 & $147143 \pm 383$ & 442\\ \hline 
    3-10 TeV & 25135 & $25021 \pm 158$ &  341 \\
  \hline       
    \end{tabular}
    \caption{Results of FRB-neutrino searches within an angular window  of $3^{\circ}$, between the FRB and neutrino directions. The signal consists of the total number of coincident  pairs within $3^{\circ}$. The  background is  obtained by averaging the data  shown in Fig.~\ref{fig3} for the off-source bins (non-shaded), equally spaced in $\cos(\theta)$ at angular separations greater than $3^{\circ}$. The error in the background represents the Poisson error.
    For all the \rthis{reconstructed muon} energy intervals, the expected number of signal events is smaller or comparable  to the background, and therefore there is no statistically significant   association between the CHIME FRBs and IceCube neutrinos. The last column contains the 90\% credible interval flux limits calculated using Eq.~\ref{eq:upperlimit}.}
    \label{tab:my_label}
\end{table}

\begin{table}[]
    \centering
    \begin{tabular}{|c|c|c|c|}
    \hline
    \textbf{Muon Energy range} & \textbf{Signal}  & \textbf{Background} & \textbf{90\% credible flux limit} \\ \hline
   Full dataset & 52965 & $53203 \pm 231$  & 262\\ \hline
    0.1 - 3 TeV & 49000 & $49322 \pm 222$  & 220 \\ \hline
    0.3 - 1 TeV & 26063 & $26327 \pm 162$  & 152 \\ \hline
    3 - 10 TeV & 3363 & $3283 \pm 57$ & 158 \\ \hline
    $>$ 10 TeV & 453 & $466 \pm 22$ & 29\\ \hline
    \end{tabular}
    \caption{Results of searches within the neutrino error region. The signal consists of the total number of coincident  pairs within the neutrino error region.
    The background is obtained by counting the number of matches within a $\cos(\delta \theta)$ angular cut, modulo a  fixed  offset of 5 degrees, where  $\delta \theta$ is the error in the neutrino position. The error in the background represents the Poisson error. Again, the number of signal events is comparable or smaller than the background for all the neutrino energy ranges analyzed. Therefore, we do not find any statistically significant association between the CHIME FRBs and IceCube neutrinos. The last column contains the 90\% credible interval flux limits calculated using Eq.~\ref{eq:upperlimit}.}
    \label{table2}
    \end{table}

\iffalse
\begin{figure}
\centering
    \includegraphics[scale=0.7]{angconematches.pdf}
    \caption{The angular separation between CHIME FRBs and IceCube neutrinos, within $30^{\circ}$. The yellow shaded regions shows the location of the bin ($0-3^{\circ}$) where one would see an excess in case of a statistically significant association between the 
    FRBs and neutrinos.}
\label{fig2}.
\end{figure}
\fi

\begin{figure}
\centering
    \includegraphics[scale=0.7]{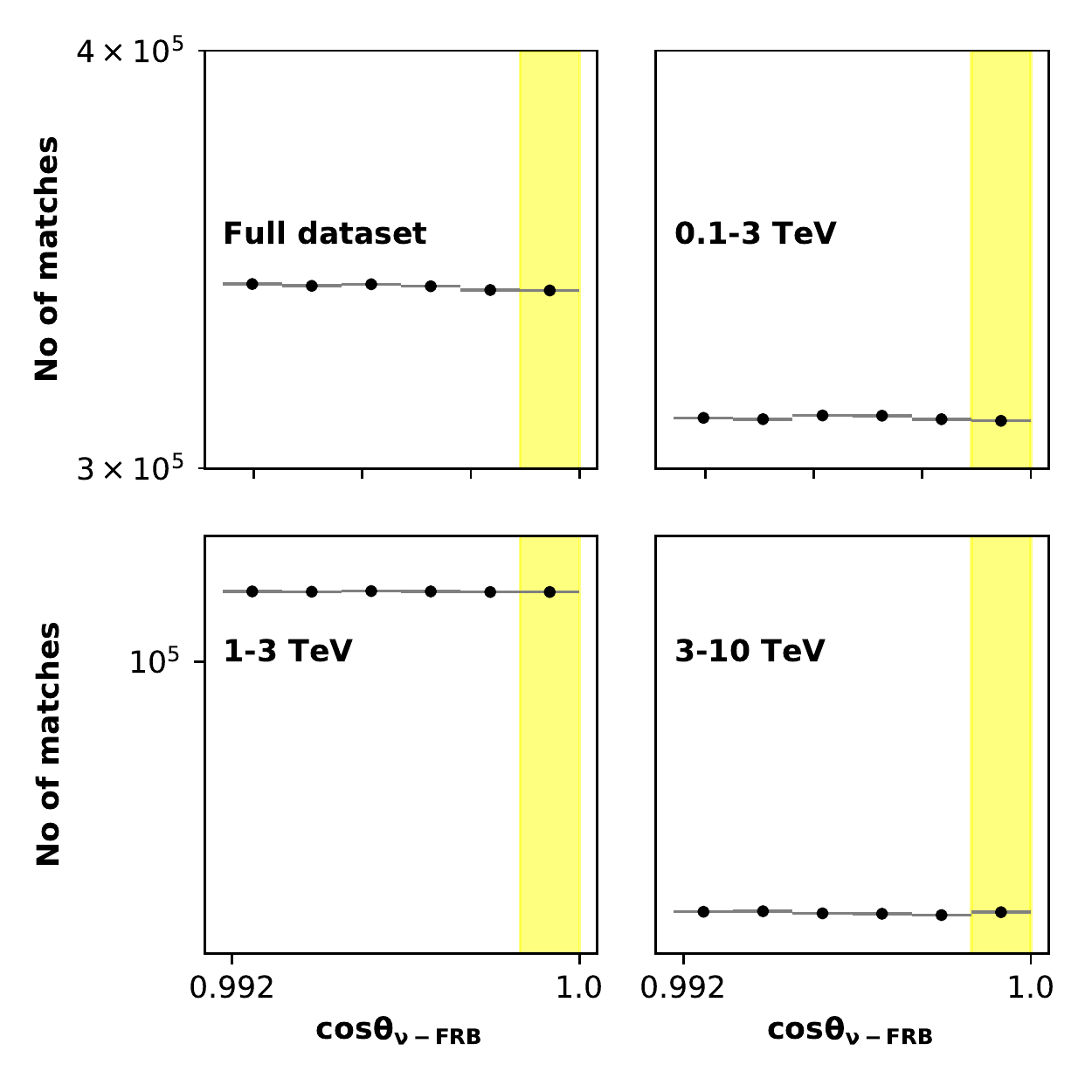}
    \caption{Angular separation between the CHIME FRBs and neutrinos in six evenly spaced  $\cos \theta$ bins from 0 to $7.35^{\circ}$. The yellow shaded region corresponds to the signal window corresponding to an  angle of $3^{\circ}$. The remaining five bins are used for background estimation. A tabular summary  comparing the signal and background can be found in Table~\ref{tab:my_label}.}
\label{fig3}.
\end{figure}

\subsection{Coincidences within neutrino error region}
It is possible that  the $3^{\circ}$ angle used for the signal window in the previous sub-section  is  too large, since the average error for the IceCube neutrinos is about $0.9^{\circ}$.  This would imply that the $3^{\circ}$ angular window is too large and  contains significant number background events. 

Therefore,  as a slight variant of the earlier analysis, we now  search for neutrino-FRB matches within the neutrino error circle, which is different for every neutrino.   To estimate the background for such a search, we count the number of neutrino-FRB pairs, whose cosine of  the angular correlation is within the cosine of the neutrino error circle,  modulo an offset of $5^{\circ}$.
In other words, if a given neutrino has an error of $\delta \theta$ and the observed angular separation between the neutrino and FRB is denoted by $\theta_{{\nu}-FRB}$, a neutrino-FRB coincident pair is considered as background, if it satisfies the following condition
\begin{equation}
|\cos (\theta_{{\nu}-FRB}) -\cos (5^{\circ})| < 0.5  [1-\cos(\delta \theta)]
\end{equation}
The factor of 0.5  ensures that the background is estimated within the same solid angle as the signal.
Any such pairs at angular separations greater than $5^{\circ}$  are due to chance coincidences, and hence can be considered as background for this search. A tabulated summary of the results from this search  for different neutrino energy ranges can be found in Table~\ref{table2}. The error in the background corresponds to the Poissonian error.
As we can see, even with this method, the observed number of coincidences within the neutrino error circle is smaller or comparable than the background events. Only for neutrinos in the 3-10 TeV energy range, we find that the total number of signal events is greater than the background. However, in this energy range, the number of excess events (80) is comparable to the error in the background (57), \rthis{corresponding to an excess of only 1.4$\sigma$. We should have observed at least 3454 matches in order to a assert a $3\sigma$ detection.}
Therefore, again we do not find any evidence for any significant angular correlation,  between the FRBs and neutrinos, by looking for matches within the neutrino error circle. We again report the 90\% credible interval upper limits on FRB-induced neutrinos from this search (using Eq.~\ref{eq:upperlimit}) in Table~\ref{tab:my_label}. In all the previous searches for neutrinos from FRBs using IceCube, both spatial and temporal searches from individual FRBs were carried out and fluence limits set on each of these FRBs. This is contrast to our analyses where our flux limits are from a stacked time-integrated searches. Therefore, a direct comparison with the previous  fluence limits  in literature is not possible.

Hence, to summarize,  we do not find any statistically significant angular  coincidence  between the CHIME FRBs and IceCube neutrinos with either of  the methods. 
%We note that one could also estimate the background by scrambling the observation times and detector coordinates and recomputing the right ascension and declination, similar to the searches done in Super-K~\cite{SKnuastro}. We shall defer this to a future work. 

\section{Conclusions}
\label{sec:conclusions}

In this work, we have searched for a spatial coincidence between IceCube neutrinos in the TeV energy range and  FRBs detected by the CHIME telescope, using background estimation methods, which were used for neutrino astrophysics searches within Super-Kamiokande.

For our analysis, we consider the IceCube neutrinos with error region $< 3^{\circ}$. We do two  searches. In the first search, we look for neutrino-FRB coincident pairs within $3^{\circ}$. The background for this search was estimated by averaging the coincident pairs in five angular bins between 3 and 7.35$^{\circ}$, where each bin has the same  width in $\cos(\theta)$, \rthis{which is  the angular size of the window used to search for signal from the FRB, viz $3^{\circ}$}. For our second search, we look for coincidences within the neutrino error circle.  For the second  search, we count the number of neutrino-FRB pairs within the same solid angle as the neutrino error circle, but offset by $5^{\circ}$, in order to emulate the backgrounds.

Our results from both the searches comparing the observed signal and background events  are summarized in Table~\ref{tab:my_label} and Table~\ref{table2}. 
For both these searches, the number of signal events is comparable to the background. Therefore, we conclude that the observed angular coincidences are consistent with background.  For both these searches, we then calculate the 90\% credible interval upper limit on the FRB-induced neutrinos,
which can be found in the aforementioned tables.

%We  should however point out that our analysis of the background estimation is simplistic. It does not incorporate the detector effective area, livetime, detection efficiency  and  other detector related effects.  These effects can be incorporated by  using the  maximum likelihood analysis,  discussed in ~\cite{Icecube10,Montaruli} . However, even this simplistic analysis is sensitive enough to detect a statistically significant  spatial correlation. Also, as  FRBs are transient sources, one should also supplement this analysis with a temporal search between the FRBs and neutrinos, to see if considering the burst or FRB arrival time would enhance the detection significance.

Note that if  FRBs are steady state point sources of astrophysical neutrinos in the TeV energy range, their detection significance should increase with additional data.  Many other radio telescopes such as uGMRT~\cite{Marthi}, FAST~\cite{Li}, VLITE~\cite{Bethapudi}  continue  to discover new FRBs and with the advent of SKA~\cite{SKA} and next generation neutrino telescopes such as IceCube Gen2~\cite{gen2}, one should be able to get a clearer picture as to whether FRBs produce  neutrinos in the TeV energy range. \rthis{In a future work, we shall also carry out temporal searches to supplement the spatial searches implemented in this work,  once the public IceCube neutrino catalog is updated.}

\section*{ACKNOWLEDGEMENT}
We are grateful to the CHIME and IceCube Collaborations for making their data publicly available and for thorough documentation of their data products. This work was originally  motivated by the preprint arXiv:2112.11375, but has been modified, following  the retraction of the preprint by the authors. We also acknowledge encouraging words and feedback from Ranjan Laha, Jonathan Katz, Clancy James, and Zorawar Wadiasingh,  after  the manuscript was posted on arXiv.

\bibliography{references}

\begin{thebibliography}{30}
\expandafter\ifx\csname natexlab\endcsname\relax\def\natexlab#1{#1}\fi
\expandafter\ifx\csname bibnamefont\endcsname\relax
  \def\bibnamefont#1{#1}\fi
\expandafter\ifx\csname bibfnamefont\endcsname\relax
  \def\bibfnamefont#1{#1}\fi
\expandafter\ifx\csname citenamefont\endcsname\relax
  \def\citenamefont#1{#1}\fi
\expandafter\ifx\csname url\endcsname\relax
  \def\url#1{\texttt{#1}}\fi
\expandafter\ifx\csname urlprefix\endcsname\relax\def\urlprefix{URL }\fi
\providecommand{\bibinfo}[2]{#2}
\providecommand{\eprint}[2][]{\url{#2}}

\bibitem[{\citenamefont{Amiri et~al.}(2021)}]{Chime}
\bibinfo{author}{\bibfnamefont{M.}~\bibnamefont{Amiri}} \bibnamefont{et~al.}
  (\bibinfo{collaboration}{CHIME/FRB}), \bibinfo{journal}{Astrophys. J. Supp.}
  \textbf{\bibinfo{volume}{257}}, \bibinfo{pages}{59} (\bibinfo{year}{2021}),
  \eprint{2106.04352}.

\bibitem[{\citenamefont{Aartsen et~al.}(2020{\natexlab{a}})}]{icecube}
\bibinfo{author}{\bibfnamefont{M.~G.} \bibnamefont{Aartsen}}
  \bibnamefont{et~al.} (\bibinfo{collaboration}{IceCube}),
  \bibinfo{journal}{Phys. Rev. Lett.} \textbf{\bibinfo{volume}{124}},
  \bibinfo{pages}{051103} (\bibinfo{year}{2020}{\natexlab{a}}),
  \eprint{1910.08488}.

\bibitem[{\citenamefont{{Lorimer} et~al.}(2007)\citenamefont{{Lorimer},
  {Bailes}, {McLaughlin}, {Narkevic}, and {Crawford}}}]{Lorimer07}
\bibinfo{author}{\bibfnamefont{D.~R.} \bibnamefont{{Lorimer}}},
  \bibinfo{author}{\bibfnamefont{M.}~\bibnamefont{{Bailes}}},
  \bibinfo{author}{\bibfnamefont{M.~A.} \bibnamefont{{McLaughlin}}},
  \bibinfo{author}{\bibfnamefont{D.~J.} \bibnamefont{{Narkevic}}},
  \bibnamefont{and}
  \bibinfo{author}{\bibfnamefont{F.}~\bibnamefont{{Crawford}}},
  \bibinfo{journal}{Science} \textbf{\bibinfo{volume}{318}},
  \bibinfo{pages}{777} (\bibinfo{year}{2007}), \eprint{0709.4301}.

\bibitem[{\citenamefont{{Petroff} et~al.}(2022)\citenamefont{{Petroff},
  {Hessels}, and {Lorimer}}}]{Petroff21}
\bibinfo{author}{\bibfnamefont{E.}~\bibnamefont{{Petroff}}},
  \bibinfo{author}{\bibfnamefont{J.~W.~T.} \bibnamefont{{Hessels}}},
  \bibnamefont{and} \bibinfo{author}{\bibfnamefont{D.~R.}
  \bibnamefont{{Lorimer}}}, \bibinfo{journal}{\aapr}
  \textbf{\bibinfo{volume}{30}}, \bibinfo{eid}{2} (\bibinfo{year}{2022}),
  \eprint{2107.10113}.

\bibitem[{\citenamefont{{Platts} et~al.}(2019)\citenamefont{{Platts},
  {Weltman}, {Walters}, {Tendulkar}, {Gordin}, and {Kandhai}}}]{Platts}
\bibinfo{author}{\bibfnamefont{E.}~\bibnamefont{{Platts}}},
  \bibinfo{author}{\bibfnamefont{A.}~\bibnamefont{{Weltman}}},
  \bibinfo{author}{\bibfnamefont{A.}~\bibnamefont{{Walters}}},
  \bibinfo{author}{\bibfnamefont{S.~P.} \bibnamefont{{Tendulkar}}},
  \bibinfo{author}{\bibfnamefont{J.~E.~B.} \bibnamefont{{Gordin}}},
  \bibnamefont{and}
  \bibinfo{author}{\bibfnamefont{S.}~\bibnamefont{{Kandhai}}},
  \bibinfo{journal}{\physrep} \textbf{\bibinfo{volume}{821}},
  \bibinfo{pages}{1} (\bibinfo{year}{2019}), \eprint{1810.05836}.

\bibitem[{\citenamefont{Andersen et~al.}(2020)}]{CHIMEmagnetar}
\bibinfo{author}{\bibfnamefont{B.~C.} \bibnamefont{Andersen}}
  \bibnamefont{et~al.} (\bibinfo{collaboration}{CHIME/FRB}),
  \bibinfo{journal}{Nature} \textbf{\bibinfo{volume}{587}}, \bibinfo{pages}{54}
  (\bibinfo{year}{2020}), \eprint{2005.10324}.

\bibitem[{\citenamefont{Aartsen et~al.}(2013{\natexlab{a}})}]{IceCube13}
\bibinfo{author}{\bibfnamefont{M.~G.} \bibnamefont{Aartsen}}
  \bibnamefont{et~al.} (\bibinfo{collaboration}{IceCube}),
  \bibinfo{journal}{Phys. Rev. Lett.} \textbf{\bibinfo{volume}{111}},
  \bibinfo{pages}{021103} (\bibinfo{year}{2013}{\natexlab{a}}),
  \eprint{1304.5356}.

\bibitem[{\citenamefont{Aartsen et~al.}(2013{\natexlab{b}})}]{IceCubescience}
\bibinfo{author}{\bibfnamefont{M.~G.} \bibnamefont{Aartsen}}
  \bibnamefont{et~al.} (\bibinfo{collaboration}{IceCube}),
  \bibinfo{journal}{Science} \textbf{\bibinfo{volume}{342}},
  \bibinfo{pages}{1242856} (\bibinfo{year}{2013}{\natexlab{b}}),
  \eprint{1311.5238}.

\bibitem[{\citenamefont{Aartsen et~al.}(2014)}]{IceCube14}
\bibinfo{author}{\bibfnamefont{M.~G.} \bibnamefont{Aartsen}}
  \bibnamefont{et~al.} (\bibinfo{collaboration}{IceCube}),
  \bibinfo{journal}{Phys. Rev. Lett.} \textbf{\bibinfo{volume}{113}},
  \bibinfo{pages}{101101} (\bibinfo{year}{2014}), \eprint{1405.5303}.

\bibitem[{\citenamefont{{M{\'e}sz{\'a}ros}}(2017)}]{Meszaros17}
\bibinfo{author}{\bibfnamefont{P.}~\bibnamefont{{M{\'e}sz{\'a}ros}}},
  \bibinfo{journal}{Annual Review of Nuclear and Particle Science}
  \textbf{\bibinfo{volume}{67}}, \bibinfo{pages}{45} (\bibinfo{year}{2017}),
  \eprint{1708.03577}.

\bibitem[{\citenamefont{{Ahlers} and {Halzen}}(2018)}]{Halzen}
\bibinfo{author}{\bibfnamefont{M.}~\bibnamefont{{Ahlers}}} \bibnamefont{and}
  \bibinfo{author}{\bibfnamefont{F.}~\bibnamefont{{Halzen}}},
  \bibinfo{journal}{Progress in Particle and Nuclear Physics}
  \textbf{\bibinfo{volume}{102}}, \bibinfo{pages}{73} (\bibinfo{year}{2018}),
  \eprint{1805.11112}.

\bibitem[{\citenamefont{{Fahey} et~al.}(2017)\citenamefont{{Fahey},
  {Kheirandish}, {Vandenbroucke}, and {Xu}}}]{Fahey}
\bibinfo{author}{\bibfnamefont{S.}~\bibnamefont{{Fahey}}},
  \bibinfo{author}{\bibfnamefont{A.}~\bibnamefont{{Kheirandish}}},
  \bibinfo{author}{\bibfnamefont{J.}~\bibnamefont{{Vandenbroucke}}},
  \bibnamefont{and} \bibinfo{author}{\bibfnamefont{D.}~\bibnamefont{{Xu}}},
  \bibinfo{journal}{\apj} \textbf{\bibinfo{volume}{845}}, \bibinfo{eid}{14}
  (\bibinfo{year}{2017}), \eprint{1611.03062}.

\bibitem[{\citenamefont{Aartsen et~al.}(2018)}]{icecube2018}
\bibinfo{author}{\bibfnamefont{M.~G.} \bibnamefont{Aartsen}}
  \bibnamefont{et~al.} (\bibinfo{collaboration}{IceCube}),
  \bibinfo{journal}{Astrophys. J.} \textbf{\bibinfo{volume}{857}},
  \bibinfo{pages}{117} (\bibinfo{year}{2018}), \eprint{1712.06277}.

\bibitem[{\citenamefont{Aartsen et~al.}(2020{\natexlab{b}})}]{IcecubeFRB}
\bibinfo{author}{\bibfnamefont{M.~G.} \bibnamefont{Aartsen}}
  \bibnamefont{et~al.} (\bibinfo{collaboration}{IceCube}),
  \bibinfo{journal}{Astrophys. J.} \textbf{\bibinfo{volume}{890}},
  \bibinfo{pages}{111} (\bibinfo{year}{2020}{\natexlab{b}}),
  \eprint{1908.09997}.

\bibitem[{\citenamefont{{Ghadimi} and {Santander}}(2021)}]{Icecubemagnetar}
\bibinfo{author}{\bibfnamefont{A.}~\bibnamefont{{Ghadimi}}} \bibnamefont{and}
  \bibinfo{author}{\bibfnamefont{M.}~\bibnamefont{{Santander}}},
  \bibinfo{journal}{arXiv e-prints} \bibinfo{eid}{arXiv:2107.08322}
  (\bibinfo{year}{2021}), \eprint{2107.08322}.

\bibitem[{\citenamefont{Abe et~al.}(2006)}]{sknuastro}
\bibinfo{author}{\bibfnamefont{K.}~\bibnamefont{Abe}} \bibnamefont{et~al.}
  (\bibinfo{collaboration}{Super-Kamiokande}), \bibinfo{journal}{Astrophys. J.}
  \textbf{\bibinfo{volume}{652}}, \bibinfo{pages}{198} (\bibinfo{year}{2006}),
  \eprint{astro-ph/0606413}.

\bibitem[{\citenamefont{Desai et~al.}(2008)}]{showering}
\bibinfo{author}{\bibfnamefont{S.}~\bibnamefont{Desai}} \bibnamefont{et~al.}
  (\bibinfo{collaboration}{Super-Kamiokande}), \bibinfo{journal}{Astropart.
  Phys.} \textbf{\bibinfo{volume}{29}}, \bibinfo{pages}{42}
  (\bibinfo{year}{2008}), \eprint{0711.0053}.

\bibitem[{\citenamefont{{Desai}}(2004)}]{Desaithesis}
\bibinfo{author}{\bibfnamefont{S.}~\bibnamefont{{Desai}}}, Ph.D. thesis,
  \bibinfo{school}{Boston University} (\bibinfo{year}{2004}).

\bibitem[{\citenamefont{Fukuda et~al.}(2002)}]{superkgrb}
\bibinfo{author}{\bibfnamefont{S.}~\bibnamefont{Fukuda}} \bibnamefont{et~al.}
  (\bibinfo{collaboration}{Super-Kamiokande}), \bibinfo{journal}{Astrophys. J.}
  \textbf{\bibinfo{volume}{578}}, \bibinfo{pages}{317} (\bibinfo{year}{2002}),
  \eprint{astro-ph/0205304}.

\bibitem[{\citenamefont{Desai et~al.}(2004)}]{superkwimp}
\bibinfo{author}{\bibfnamefont{S.}~\bibnamefont{Desai}} \bibnamefont{et~al.}
  (\bibinfo{collaboration}{Super-Kamiokande}), \bibinfo{journal}{Phys. Rev. D}
  \textbf{\bibinfo{volume}{70}}, \bibinfo{pages}{083523}
  (\bibinfo{year}{2004}), \eprint{hep-ex/0404025}.

\bibitem[{\citenamefont{{Li} et~al.}(2014)\citenamefont{{Li}, {Zhou}, {He},
  {Fan}, and {Wei}}}]{Li14}
\bibinfo{author}{\bibfnamefont{X.}~\bibnamefont{{Li}}},
  \bibinfo{author}{\bibfnamefont{B.}~\bibnamefont{{Zhou}}},
  \bibinfo{author}{\bibfnamefont{H.-N.} \bibnamefont{{He}}},
  \bibinfo{author}{\bibfnamefont{Y.-Z.} \bibnamefont{{Fan}}}, \bibnamefont{and}
  \bibinfo{author}{\bibfnamefont{D.-M.} \bibnamefont{{Wei}}},
  \bibinfo{journal}{\apj} \textbf{\bibinfo{volume}{797}}, \bibinfo{eid}{33}
  (\bibinfo{year}{2014}), \eprint{1312.5637}.

\bibitem[{\citenamefont{{Gupta} and {Saini}}(2018)}]{Patrick}
\bibinfo{author}{\bibfnamefont{P.~D.} \bibnamefont{{Gupta}}} \bibnamefont{and}
  \bibinfo{author}{\bibfnamefont{N.}~\bibnamefont{{Saini}}},
  \bibinfo{journal}{Journal of Astrophysics and Astronomy}
  \textbf{\bibinfo{volume}{39}}, \bibinfo{eid}{14} (\bibinfo{year}{2018}),
  \eprint{1709.00185}.

\bibitem[{\citenamefont{Abbasi et~al.}(2021)}]{icecubedata}
\bibinfo{author}{\bibfnamefont{R.}~\bibnamefont{Abbasi}} \bibnamefont{et~al.}
  (\bibinfo{collaboration}{IceCube}) (\bibinfo{year}{2021}),
  \eprint{2101.09836}.

\bibitem[{\citenamefont{Abe et~al.}(2016)}]{sksolar}
\bibinfo{author}{\bibfnamefont{K.}~\bibnamefont{Abe}} \bibnamefont{et~al.}
  (\bibinfo{collaboration}{Super-Kamiokande}), \bibinfo{journal}{Phys. Rev. D}
  \textbf{\bibinfo{volume}{94}}, \bibinfo{pages}{052010}
  (\bibinfo{year}{2016}), \eprint{1606.07538}.

\bibitem[{\citenamefont{Zyla et~al.}(2020)}]{PDG}
\bibinfo{author}{\bibfnamefont{P.}~\bibnamefont{Zyla}} \bibnamefont{et~al.}
  (\bibinfo{collaboration}{Particle Data Group}), \bibinfo{journal}{PTEP}
  \textbf{\bibinfo{volume}{2020}}, \bibinfo{pages}{083C01}
  (\bibinfo{year}{2020}), \bibinfo{note}{and 2021 update}.

\bibitem[{\citenamefont{{Marthi} et~al.}(2022)\citenamefont{{Marthi},
  {Bethapudi}, {Main}, {Lin}, {Spitler}, {Wharton}, {Li}, {Gautam}, {Pen}, and
  {Hilmarsson}}}]{Marthi}
\bibinfo{author}{\bibfnamefont{V.~R.} \bibnamefont{{Marthi}}},
  \bibinfo{author}{\bibfnamefont{S.}~\bibnamefont{{Bethapudi}}},
  \bibinfo{author}{\bibfnamefont{R.~A.} \bibnamefont{{Main}}},
  \bibinfo{author}{\bibfnamefont{H.~H.} \bibnamefont{{Lin}}},
  \bibinfo{author}{\bibfnamefont{L.~G.} \bibnamefont{{Spitler}}},
  \bibinfo{author}{\bibfnamefont{R.~S.} \bibnamefont{{Wharton}}},
  \bibinfo{author}{\bibfnamefont{D.~Z.} \bibnamefont{{Li}}},
  \bibinfo{author}{\bibfnamefont{T.}~\bibnamefont{{Gautam}}},
  \bibinfo{author}{\bibfnamefont{U.~L.} \bibnamefont{{Pen}}}, \bibnamefont{and}
  \bibinfo{author}{\bibfnamefont{G.~H.} \bibnamefont{{Hilmarsson}}},
  \bibinfo{journal}{\mnras} \textbf{\bibinfo{volume}{509}},
  \bibinfo{pages}{2209} (\bibinfo{year}{2022}), \eprint{2108.00697}.

\bibitem[{\citenamefont{Li et~al.}(2021)\citenamefont{Li, Wang, Zhu, Zhang,
  Zhang, Duan, Zhang, Feng, Tang, Chatterjee et~al.}}]{Li}
\bibinfo{author}{\bibfnamefont{D.}~\bibnamefont{Li}},
  \bibinfo{author}{\bibfnamefont{P.}~\bibnamefont{Wang}},
  \bibinfo{author}{\bibfnamefont{W.}~\bibnamefont{Zhu}},
  \bibinfo{author}{\bibfnamefont{B.}~\bibnamefont{Zhang}},
  \bibinfo{author}{\bibfnamefont{X.}~\bibnamefont{Zhang}},
  \bibinfo{author}{\bibfnamefont{R.}~\bibnamefont{Duan}},
  \bibinfo{author}{\bibfnamefont{Y.}~\bibnamefont{Zhang}},
  \bibinfo{author}{\bibfnamefont{Y.}~\bibnamefont{Feng}},
  \bibinfo{author}{\bibfnamefont{N.}~\bibnamefont{Tang}},
  \bibinfo{author}{\bibfnamefont{S.}~\bibnamefont{Chatterjee}},
  \bibnamefont{et~al.}, \bibinfo{journal}{Nature}
  \textbf{\bibinfo{volume}{598}}, \bibinfo{pages}{267} (\bibinfo{year}{2021}).

\bibitem[{\citenamefont{{Bethapudi} et~al.}(2021)\citenamefont{{Bethapudi},
  {Kerr}, {Ray}, {Clarke}, {Kassim}, and {Deneva}}}]{Bethapudi}
\bibinfo{author}{\bibfnamefont{S.}~\bibnamefont{{Bethapudi}}},
  \bibinfo{author}{\bibfnamefont{M.}~\bibnamefont{{Kerr}}},
  \bibinfo{author}{\bibfnamefont{P.~S.} \bibnamefont{{Ray}}},
  \bibinfo{author}{\bibfnamefont{T.~E.} \bibnamefont{{Clarke}}},
  \bibinfo{author}{\bibfnamefont{N.~E.} \bibnamefont{{Kassim}}},
  \bibnamefont{and} \bibinfo{author}{\bibfnamefont{J.~S.}
  \bibnamefont{{Deneva}}}, \bibinfo{journal}{Research Notes of the American
  Astronomical Society} \textbf{\bibinfo{volume}{5}}, \bibinfo{eid}{46}
  (\bibinfo{year}{2021}).

\bibitem[{\citenamefont{{Hashimoto} et~al.}(2020)\citenamefont{{Hashimoto},
  {Goto}, {On}, {Lu}, {Santos}, {Ho}, {Wang}, {Kim}, and {Hsiao}}}]{SKA}
\bibinfo{author}{\bibfnamefont{T.}~\bibnamefont{{Hashimoto}}},
  \bibinfo{author}{\bibfnamefont{T.}~\bibnamefont{{Goto}}},
  \bibinfo{author}{\bibfnamefont{A.~Y.~L.} \bibnamefont{{On}}},
  \bibinfo{author}{\bibfnamefont{T.-Y.} \bibnamefont{{Lu}}},
  \bibinfo{author}{\bibfnamefont{D.~J.~D.} \bibnamefont{{Santos}}},
  \bibinfo{author}{\bibfnamefont{S.~C.~C.} \bibnamefont{{Ho}}},
  \bibinfo{author}{\bibfnamefont{T.-W.} \bibnamefont{{Wang}}},
  \bibinfo{author}{\bibfnamefont{S.~J.} \bibnamefont{{Kim}}}, \bibnamefont{and}
  \bibinfo{author}{\bibfnamefont{T.~Y.~Y.} \bibnamefont{{Hsiao}}},
  \bibinfo{journal}{\mnras} \textbf{\bibinfo{volume}{497}},
  \bibinfo{pages}{4107} (\bibinfo{year}{2020}), \eprint{2008.00007}.

\bibitem[{\citenamefont{{Aartsen} et~al.}(2021)\citenamefont{{Aartsen},
  {Abbasi}, {Ackermann}, {Adams}, {Aguilar}, {Ahlers}, {Ahrens}, {Alispach},
  {Allison}, {Amin} et~al.}}]{gen2}
\bibinfo{author}{\bibfnamefont{M.~G.} \bibnamefont{{Aartsen}}},
  \bibinfo{author}{\bibfnamefont{R.}~\bibnamefont{{Abbasi}}},
  \bibinfo{author}{\bibfnamefont{M.}~\bibnamefont{{Ackermann}}},
  \bibinfo{author}{\bibfnamefont{J.}~\bibnamefont{{Adams}}},
  \bibinfo{author}{\bibfnamefont{J.~A.} \bibnamefont{{Aguilar}}},
  \bibinfo{author}{\bibfnamefont{M.}~\bibnamefont{{Ahlers}}},
  \bibinfo{author}{\bibfnamefont{M.}~\bibnamefont{{Ahrens}}},
  \bibinfo{author}{\bibfnamefont{C.}~\bibnamefont{{Alispach}}},
  \bibinfo{author}{\bibfnamefont{P.}~\bibnamefont{{Allison}}},
  \bibinfo{author}{\bibfnamefont{N.~M.} \bibnamefont{{Amin}}},
  \bibnamefont{et~al.}, \bibinfo{journal}{Journal of Physics G Nuclear Physics}
  \textbf{\bibinfo{volume}{48}}, \bibinfo{eid}{060501} (\bibinfo{year}{2021}),
  \eprint{2008.04323}.

\end{thebibliography}
\end{document}